\documentclass[%
 reprint,
 amsmath,amssymb,
 aps,
]{revtex4-2}

\usepackage{graphicx}% Include figure files
\usepackage{dcolumn}% Align table columns on decimal point
\usepackage{bm}% bold math
\usepackage{braket}
\usepackage{comment}
\def\iu{\mathrm{i}}
\begin{document}

\title{Ultrafast dynamics of entanglement in  Heisenberg antiferromagnets}
\author{G. Fabiani} \author{J. H. Mentink}
\affiliation{Radboud University, Institute for Molecules and Materials (IMM) Heyendaalseweg 135, 6525 AJ Nijmegen, The Netherlands}
\date{\today}

\begin{abstract}
We investigate entanglement dynamics in the antiferromagnetic Heisenberg model in two dimensions following a spatially anisotropic quench of the exchange interactions. Opposed to established results in one dimension, the magnon quasiparticles show an initial growth of entanglement dynamics that does not depend on the system size and is governed by the oscillation period of the exchange interaction. We ascribe this to the dominance of the intrinsic entanglement of short wavelength non-propagating magnon-pairs, which also leads to a competition between area-law and volume-law contribution in the entanglement dynamics. Furthermore, by adopting the neural-network quantum states, we provide numerical evidence that this behavior survives even in the presence of strong magnon-magnon interactions, suggesting new avenues for manipulating entanglement dynamics in quantum materials.
\end{abstract}
\maketitle

\section{\textbf{Introduction}}
The study of entanglement in non-equilibrium quantum systems has gained considerable attention in recent years, fueled by the possibility to probe the dynamics of quantum correlations in systems such as cold atoms \cite{Greiner, Lewenstein} and trapped ions  \cite{Leibfried, Blatt}, which provide experimental access to fundamental questions concerning the onset of thermalization in isolated quantum many-body systems \cite{Eisert, Gogolin, Rigol}. 

The dynamics of entanglement has been studied in a wide variety of one-dimensional systems, following quantum quenches \cite{De Chiara, Lauchli, Fagotti, Calabrese2009, Kormos2014, Collura, Hazzard, Buyskikh, Kormos2016, Ho, Alba, Calabrese2020}.
Due to a widely accepted semi-classical approach pioneered by Calabrese and Cardy \cite{Calabrese}, the post-quench dynamics can be understood as ballistic propagation of pairs of entangled quasiparticles that, as they move, correlate spatially separated regions of the system. As a consequence, entanglement grows linearly with time, with a speed determined by the highest quasiparticle velocity, setting a fundamental bound for the time scale at which thermalization can set in.

In principle, recently developed ultrafast pump-probe techniques make it possible to investigate entanglement dynamics in extended quantum materials as well. 
However, in these systems the light-matter interaction in general breaks rotation invariance, even in the simplest electric dipole approximation,  yielding homogeneous (\emph{i.e.}  translationally invariant) yet spatially anisotropic perturbations. For this reason, it is unclear how to link the dynamics in these systems to the isotropic quenches considered so far.
Studying the entanglement dynamics in quantum materials requires, therefore, to uncover the non-equilibrium dynamics following spatially anisotropic excitation protocols and this may reveal how the quasiparticle picture generalizes to high dimensional and spatially anisotropic cases.

A prime example of spatially anisotropic control of quantum interactions is the ultrafast control of exchange interactions, which recently has emerged as a central tool for the manipulation of quantum materials \cite{Mentink2014, Mentink2015, Claassen, Kitamura, Liu2018, Chaudhary, Barbeau, Mikhaylovskiy2015, Mikhaylovskiy2020, Losada, Sriram, Ke, Sentef2020, Ono, Wang}. In this case, the light-matter interaction depends on the relative orientation of the laser field polarization and the exchange bonds, resulting in homogeneous but spatially anisotropic perturbations.  

The class of antiferromagnetic quantum materials is particularly interesting, since antiferromagnets feature magnon entanglement even in the ground state, which manifests as magnon squeezing between the two opposite sublattices \cite{Kamra2019, Wuhrer2022}. Intriguingly, similar two-magnon squeezing has been observed also in the time-domain following ultrashort anisotropic perturbations of the exchange interaction \cite{Zhao}, opening up new avenues to control and harness entanglement at the timescale of the exchange interactions. 

By focusing on the two-dimensional Heisenberg antiferromagnet, the effect of anisotropic quenches of the exchange interaction on the propagation of the excited magnons was recently analyzed in \cite{Fabiani2}. However, the dynamics of entanglement has not been investigated so far and is the main objective of this work. Interestingly, despite that this is a locally interacting system with well-defined quasiparticles, we find that the entanglement of these quasiparticles themselves yields entanglement dynamics that greatly deviates from the quasiparticle picture.

We provide both analytical and numerical evidence that the entanglement entropy increases and oscillates in a time-scale determined by the exchange interaction, reminiscent to the dynamics of the nearest neighbour spin correlations. For typical values of the exchange interactions, such dynamics is ultrafast, \emph{i.e. }in the sub-picosecond regime. We explain this by showing that the entanglement dynamics is dominated by the intrinsic entanglement of short wavelength magnon pairs, which are favored by the anisotropic quench. Furthermore, the presence of intrinsic entanglement leads to contributions of both \emph{area law} and  \emph{volume law} scaling of the entanglement dynamics with system size. By employing the recently proposed neural-network quantum states, we show that the rapid entanglement growth persists even in the presence of strong magnon-magnon interactions.

\section{\textbf{Magnon-pair dynamics}}
\begin{figure}
 \includegraphics[width=8.6cm]{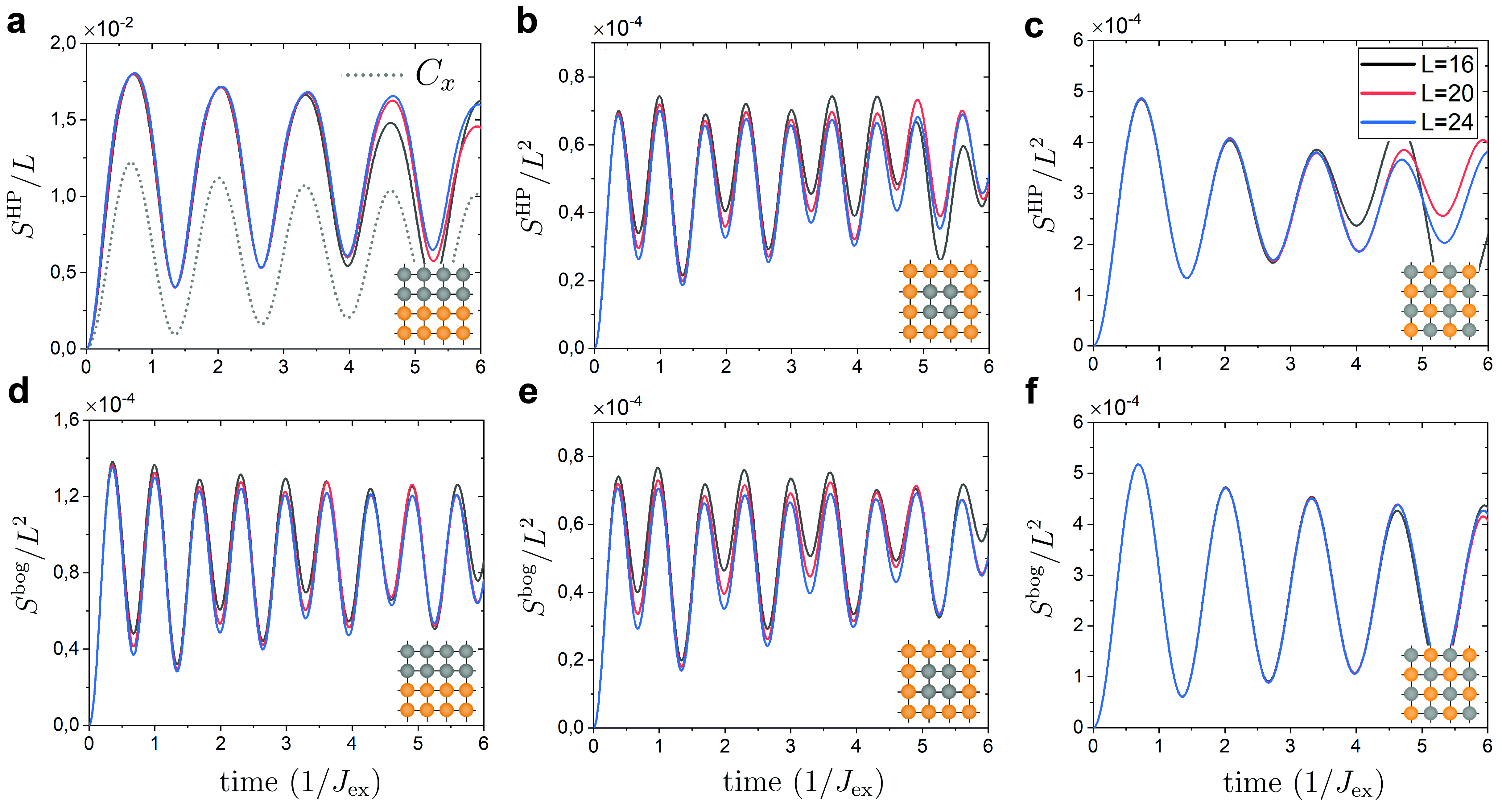}%
    \caption{(Color online) Dynamics of the entanglement entropy for three different partitions in the Holstein-Primakoff (a)-(c) and Bogoliubov basis (d)-(f) for $L=16$ (back), $L=20$ (red), $L=24$ (blue). The three partitions have respectively $L\times L/2$  (a,c), $L/2 \times L/2$ (b,d) and $L/2\times L$ spins (c,f). Dotted gray line: nearest neighbour correlations $C_x(t)$.}%
    \label{fig:S_lswt}%
  \end{figure} 
We consider the spin-$\frac{1}{2}$ antiferromagnetic Heisenberg model on a square lattice with $N=L\times L$ spins $\hat{S}_i=\hat{S}(\mathbf{r}_i)$, with $\mathbf{r}_i=(x_i,y_i)$, described by the Hamiltonian   
\begin{equation}\label{eq:1}
\hat{\mathcal{H}} = J_{\text{ex}} \sum_{\langle ij\rangle } \hat{S}_i \cdot \hat{S}_{j},
\end{equation}
where $J_{\text{ex}}$ is the exchange interaction ($J_{\textrm{ex}}>0$) and $\langle \cdot \rangle$ restricts the sum to nearest neighbours. To excite non-equilibrium spin dynamics we consider the following perturbation of the exchange interaction
\begin{equation}\label{eq:2}
\delta \hat{\mathcal{H}}(t) = \Delta J_{\textrm{ex}}(t)\sum_{i,\pmb{\delta}}\big(\mathbf{e}\cdot \pmb{\delta}\,\big)^2\hat{S}(\mathbf{r}_i) \cdot \hat{S}(\mathbf{r}_i+\pmb{\delta}),
\end{equation}
where $\mathbf{e}$ is a unit vector that determines the polarization of the electric field of the light pulse which causes the perturbation and $\pmb{\delta}$ connects nearest neighbour spins. This perturbation has been employed in several works to model the setup of impulsive stimulated Raman scattering \cite{Zhao, Bossini 2016, Fleury, Deveraux, Bossini 2019, Weber}, for which $\Delta J_\text{ex}(t)=\delta(t)$. Here instead, in order to be closer to typical setups of non-equilibrium quantum dynamics, we consider a global quench, namely we set $\Delta J_\text{ex}(t) = 0.1 J_\text{ex}\,\Theta(t)$, where $\Theta(t)$ is the Heaviside step-function. Nevertheless, as previously found \cite{Fabiani, Fabiani2}, the two protocols feature qualitatively similar dynamics. 
In the following we set $\hbar=1$, the lattice constant $a=1$ and we choose $\mathbf{e}$ along the y-axis of the square lattice.

Magnon-pair dynamics can be solved analytically treated in the linear spin wave approximation (LSWT). To this end, we perform the linear Holstein-Primakoff transformations $\hat{{S}}_{i}^z = S-\hat{a}^\dagger_{i}\hat{a}_{i}, \; \hat{{S}}^+_{i} = \sqrt{2S}\hat{a}_{i},\;  \hat{{S}}^-_{i} = \sqrt{2S}\hat{a}^\dagger_{i}$, which in Fourier space give (up to constant terms)
\begin{equation} \label{eq:H_hp}
\hat{\mathcal{H}} = \frac{zJ_{\text{ex}}S}{2}\sum_{\mathbf{k}}\bigg[\hat{a}_{\mathbf{k}}^\dagger\hat{a}_{\mathbf{k}} + \hat{a}_{\mathbf{-k}}\hat{a}^\dagger_{\mathbf{-k}} - \eta\gamma_{\mathbf{k}}\Big(\hat{a}_{\mathbf{k}}^\dagger\hat{a}_{-\mathbf{k}}^\dagger+\hat{a}_{\mathbf{k}}\hat{a}_{-\mathbf{k}}\Big)\bigg],
\end{equation}
where $z=4$ is the coordination number and $\gamma_{\mathbf{k}} = \frac{1}{z}\sum_{\bm{\delta}}e^{\mathbf{k}\cdot \bm{\delta}}$. 
In this step we have added a staggered field to Eq. \eqref{eq:1}, which results in a factor $\eta$ in front of $\gamma_{\mathbf{k}}$. The value of $\eta$ is then adjusted to guarantee that, for finite systems, the staggered magnetization is zero. This \emph{modified} spin wave theory \cite{Hirsch} restores the sublattice invariance of the Heisenberg Hamiltonian (otherwise broken in the conventional spin wave treatment), and in a previous work was found to yield accurate results for the ground state entanglement of the 2D Heisenberg model \cite{Song}. 

The Hamiltonian Eq. \eqref{eq:H_hp} can be diagonalized via a Bogoliubov transformation $\hat{\alpha}_{\mathbf{k}} = u_{\mathbf{k}} \,\hat{a}_{\mathbf{k}} - v_{\mathbf{k}} \,\hat{a}^\dagger_{-\mathbf{k}}$, yielding 
\begin{eqnarray}\label{eq:H_bog}
\hat{\mathcal{H}} + \delta \hat{\mathcal{H}}(t) = &\sum_{\mathbf{k}}& \frac{1}{2} \Big[\big(\omega_\mathbf{k} + \delta\omega_\mathbf{k}\big)\big(\hat{\alpha}^\dagger_\mathbf{k}\hat{\alpha}_\mathbf{k} + \hat{\alpha}_{-\mathbf{k}}\hat{\alpha}^\dagger_{-\mathbf{k}}\big)\nonumber\\
 &+& V_\mathbf{k}\big(\hat{\alpha}^\dagger_\mathbf{k}\hat{\alpha}^\dagger_{-\mathbf{k}}+\hat{\alpha}_\mathbf{k}\hat{\alpha}_{-\mathbf{k}}\big)\Big].
\end{eqnarray}
Here $\omega_\mathbf{k}$ is the single-magnon dispersion renormalized by the Oguchi correction \cite{Oguchi}, while  $\delta\omega_\mathbf{k}$ and $V_\mathbf{k}$ are proportional to $\Delta J_{\text{ex}}(t)$ and depend on the details of the perturbation, see Appendix \ref{App:A} for further information. The first term describes the bare magnon spectrum, which is renormalized due to the perturbation of $J_{\text{ex}}$. The second term originates only from Eq. \eqref{eq:2} and is responsible for the creation and annihilation of pairs of counter-propagating magnons. We want to emphasize that the spin-flip excitation of Eq. \eqref{eq:2} is homogeneous (\textit{i.e.} translation invariant) in the lattice space, but highly localized to nearest neighbouring bonds. This favours the excitation of high-$\mathbf{k}$ magnons, which is reflected on a hierarchy of $V_\mathbf{k}$ from a vanishing value at the center of the Brillouin zone to the largest value at the edge where $\mathbf{k}=\mathbf{X}\equiv(0,\pi)$.

\section{\textbf{Entanglement of magnon-pairs}}

\begin{figure}
  \includegraphics[width=8.6cm]{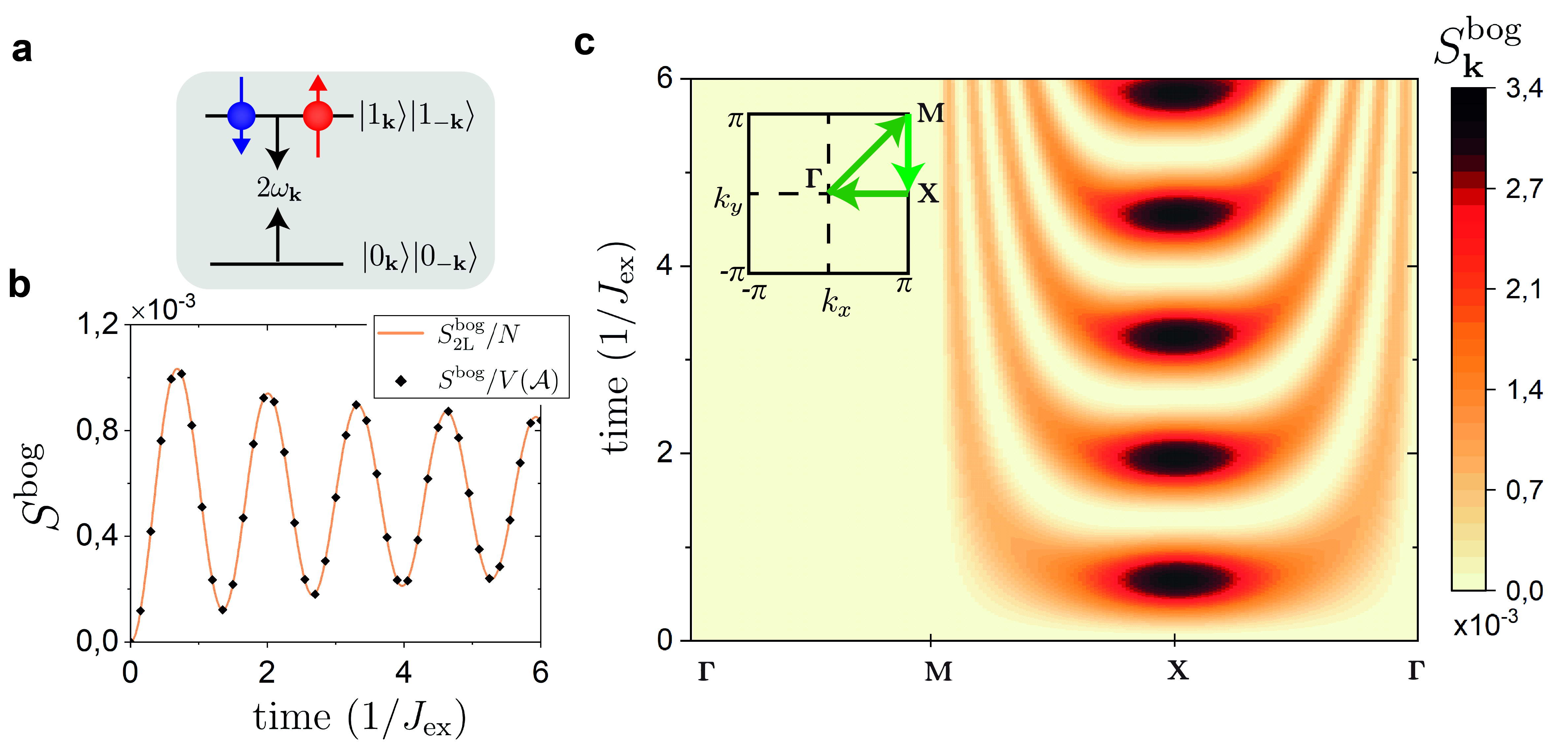}%
    \caption{(Color online) Dominance of short-wavelength magnons in the entanglement entropy. (a) Two-level system representation. (b) Dynamics of the entanglement entropy in the Bogoliubov basis for $L=24$: comparison between $S^{\text{bog}}_{\text{2L}}/N$ (orange solid line) and $S^{\text{bog}}/V(\mathcal{A})$ for the checkerboard partition with $V(\mathcal{A})=L\times L/2$ (black diamonds). (c) Time-evolution of $S_{\mathbf{k}}^{\text{bog}}$ along the symmetry line of the Brillouin zone shown in the inset (green arrows), for a system with $L=120$.}%
    \label{fig:S_lswt_bog}%
  \end{figure} 

To quantify the entanglement dynamics following the quench we employ the second order Renyi entropy
\begin{equation} \label{eq:S_2}
S_2 (\rho_{\mathcal{A}}) = -\log \big(\text{Tr} \rho_{\mathcal{A}}^2\big) = -\log \big(\text{Tr} \rho_{\mathcal{B}}^2\big)=S_2(\rho_{\mathcal{B}}),
\end{equation}
between two complementary partitions $\mathcal{A}$ and $\mathcal{B}$ of the square lattice, where $\rho_{\mathcal{A}(\mathcal{B})} = \text{Tr}_{\mathcal{B}(\mathcal{A})} \ket{\Psi}\bra{\Psi}$ and $\text{Tr}_{\mathcal{A}}$ ($\text{Tr}_{\mathcal{B}}$) is the trace of the $\mathcal{A}$ ($\mathcal{B}$) degrees of freedom in the Hilbert space $\mathcal{H}=\mathcal{H}_{\mathcal{A}}\otimes \mathcal{H}_{\mathcal{B}}$.
In the linear spin wave approximation the Hamiltonian is quadratic and this allows to express the reduced density matrices solely in terms of single-particle correlation functions, yielding a simple procedure to evaluate $S_2(\rho_{\mathcal{A}})$ \cite{Peschel}. Specifically, we define the matrices $X_{ij} = f(\mathbf{r}_i - \mathbf{r}_j) + g(\mathbf{r}_i - \mathbf{r}_j)$ and $P_{ij} = f(\mathbf{r}_i - \mathbf{r}_j) - g(\mathbf{r}_i - \mathbf{r}_j)$, with
\begin{eqnarray}\label{eq:f}
f(\mathbf{r}_i - \mathbf{r}_j) &=& \frac{1}{2}\braket{\hat{b}^\dagger_i\hat{b}_j} + \braket{\hat{b}_i\hat{b}^\dagger_j}, \\ \label{eq:g}
g(\mathbf{r}_i - \mathbf{r}_j) &=& \frac{1}{2}\braket{\hat{b}^\dagger_i\hat{b}^\dagger_j} + \braket{\hat{b}_i\hat{b}_j},
\end{eqnarray}
where $\mathbf{r}_{i,j} \in \mathcal{A(B)}$ and $\hat{b}_i,\,\hat{b}_i^\dagger$ are the annihilation/creation operators of a chosen basis. If $\zeta_p$ are the eigenvalues of the matrix $Q(\mathbf{r}_i - \mathbf{r}_j) = \big[X\cdot P \big]_{ij}$, then the Renyi entropy can be evaluated from
\begin{equation}\label{eq:S_LSWT}
S_2(\rho_{\mathcal{A(B)}}) = \sum_p \ln 2\zeta_p,
\end{equation}
and $\zeta_p \geq 1/2$. While $f(\mathbf{r}_i - \mathbf{r}_j)$ and $g(\mathbf{r}_i - \mathbf{r}_j)$ can be evaluated analytically, the eigenvalues of $Q$ have to be evaluated numerically for the system sizes considered.
In the following we are going to calculate the Renyi entropy for three different partitions of the square lattice. This is done both in the Holstein-Primakoff and Bogoliubov basis by choosing the corresponding bosons in Eqs. \eqref{eq:f}-\eqref{eq:g}. The former is closely related to the entanglement in the spin basis of Eqs. \eqref{eq:1}-\eqref{eq:2}, while the latter measures the magnon entanglement for which it is possible to get an analytical expression for $S_2(\rho)$, that will be used to interpret the results.

The dynamics of the entanglement entropy in the Holstein-Primakoff basis $S^{\text{HP}}(t)= S^{\text{HP}}_2\big(\rho_{\mathcal{A}}(t)\big)-S^{\text{HP}}_2\big(\rho_{\mathcal{A}}(0)\big)$ is shown in Figs. \ref{fig:S_lswt}(a)-(c) for three different partitions of the lattice and $L=16,\,20,\,24$. The striking feature that emerges is that the entanglement dynamics greatly deviates from the evolution expected by a straightforward application of the quasiparticle picture, which predicts an entanglement growth in the time-scale of $L/(2\mathbf{v}_{\text{max}})$, where $\mathbf{v}_{\text{max}} = \frac{d\omega_{\mathbf{k}}}{d\mathbf{k}}|_{\mathbf{k}=0}\sim 1.64\,J_{\text{ex}}$ corresponds to the highest magnon group velocity. Here, instead, the initial growth happens in a time-window independent of system size and is followed by periodic oscillations at a frequency only determined by the exchange interaction. In particular, for the partition shown in Fig. \ref{fig:S_lswt}(a), the dynamics of entanglement resembles the oscillations of the nearest neighbour spin correlations $C_x(t) = \braket{\hat{S}(\mathbf{r}_i)(t)\cdot \hat{S}(\mathbf{r}_i + \pmb{\delta}_x)(t)}$ (grey dotted line), while for the other partitions $S^{\text{HP}}$ oscillates at twice the frequency of $C_x(t)$. This suggests that the short wavelength magnons at the edge of the Brillouin zone, which dominantly contribute to the dynamics of $C_x(t)$, play a prominent role in the time-evolution of the entanglement entropy.

To further support the origin of this rapid entanglement dynamics and the relation with high-$\mathbf{k}$ magnons, we next consider the entanglement directly from the Bogoliubov bosons. In this case the initial state is the vacuum state and has therefore no entanglement. This setting is therefore closer to conventional quench scenarios in one dimension. Nevertheless, the dynamics of the entanglement entropy $S^{\text{bog}}(t)$ (Fig. \ref{fig:S_lswt}(d, f)) still deviates from the quasiparticle picture and closely resembles the corresponding dynamics in the HP basis.
Importantly, in this basis an analytical expression can be obtained for the \emph{checkerboard} partition of Fig. \ref{fig:S_lswt}(f). This relies on the fact that the counter-propagating magnons of each excited pair originate from opposite sublattices corresponding to sublattice $\mathcal{A}$ and $\mathcal{B}$ of the checkerboard partition \cite{Zhao, Bossini 2019} and the reduced density matrix for one sublattice can be obtained by tracing out the degrees of freedom of one of the two bosons. This can be easily done by restricting the Hilbert space to a collection of independent two-level systems, each of them encompassing the vacuum state and a state with a pair of magnons \cite{Mentink2017}. The wavefunction of the entire system can then be written as $\ket{\psi(t)} = \prod_{\mathbf{k}}\ket{\psi_{\mathbf{k}}(t)}$, with $\ket{\psi_{\mathbf{k}}(t)} = c_{\mathbf{k}}(t) \ket{0_{\mathbf{k}} 0_{-\mathbf{k}}} + d_{\mathbf{k}}(t) \ket{1_{\mathbf{k}} 1_{-\mathbf{k}}}$. The time-dependent coefficients $c_{\mathbf{k}}(t)$ and $d_{\mathbf{k}}(t)$ are found by solving the Schroedinger equation $i\partial_t \ket{\psi(t)} = (\hat{H} + \delta\hat{H})\ket{\psi(t)}$ with the initial conditions $c_{\mathbf{k}}(0)=1$ and $d_{\mathbf{k}}(0)=0$ (see Appendix \ref{App:B}). With this ansatz, the second order Renyi entropy between the two modes of each two-level system reads 
\begin{equation}\label{eq:S_bog}
S_{\mathbf{k}}^{\text{bog}}(t) =- \ln \big(|c_{\mathbf{k}}(t)|^4 + |d_{\mathbf{k}}(t)|^4\big),
\end{equation}
and the entanglement of the entire system is given by $S^{\text{bog}}_{\text{2L}}(t) = \sum_{\mathbf{k}} S_{\mathbf{k}}^{\text{bog}}(t)$, where "2L" remarks that the Renyi entropy is calculated in the two-level system approximation. The dynamics of $S^{\text{bog}}_{\text{2L}}(t)$ is shown in Fig. \ref{fig:S_lswt_bog}(b) and we find perfect overlap with the calculation performed by using Eq. \eqref{eq:S_LSWT} with Bogoliubov bosons.  However, we can now employ Eq. \eqref{eq:S_bog} to get an analytic understanding of the entanglement dynamics. In particular, we note that at each time, the largest terms in the sum are those given by the zone-edge magnons with $\mathbf{k}\approx\mathbf{X}$ (see Fig. \ref{fig:S_lswt_bog}(c)), which follows from the dominance of $V_\mathbf{k}$ in the vicinity of the zone-edge. Moreover, due to a Van Hove singularity in the magnon density of states at the zone edge, the number of short wavelength magnons greatly exceeds that of long wavelength magnons. As a result, the entanglement dynamics is fully dominated by the oscillations of zone-edge magnon entanglement and can be approximated as $S^{\text{bog}}_{\text{2L}}(t) \approx D_{\mathbf{X}} \ln \big(|c_{\mathbf{X}}|^4 + |d_{\mathbf{X}}|^4\big)$, where $D_{\mathbf{X}}$ is the amplitude of the magnon density of states at ${\mathbf{k}}\approx{\mathbf{X}}$. Similar dynamics is found as well in Fig.~\ref{fig:S_lswt}(c), since for these wave vectors the Bogoliubov transformation is nearly the identity \cite{Bossini 2019}.
Pairs of long wavelength magnons, which determine the light-cone dynamics of correlations, also contribute to the growth of entanglement, but their contribution is eclipsed by the intrinsic entanglement dynamics of each high-energy magnon pair. Therefore, our results are not in contradiction with the quasiparticle picture, but reveal a new scenario where spreading of correlations and growth of entanglement originate from excitations with different energy scales. 

Interestingly, we find that also the scaling of entanglement is altered by the localized character of the spin-flip excitation considered. In Fig. \ref{fig:S_lswt}(a) we observe that the entanglement dynamics follows an area law, while a volume law is found in the other cases. This can be qualitatively explained as follows: From the analytical calculation above, we expect the entanglement to scale with the number of magnons in close vicinity to the zone-edge and this increases with the system size, resulting in a volume law entanglement dynamics. At the same time, the dominance of high-$\mathbf{k}$ magnons causes exceptionally large nearest neighbour correlations in the spin basis, which suggests an important area law contribution for $S^{\text{HP}}$. Due to the symmetry of the perturbation \cite{Fabiani2}, the nearest neighbour correlations along orthogonal directions of the lattice oscillate out of phase and therefore they yield opposite contributions to $S_2(\rho_{\mathcal{A}})$ (see Appendix \ref{App:C}; similar results are also found in photo-doped Mott insulators \cite{Tsutsui}). The area law contribution from boundary correlations vanishes for partitions with equal number of correlations along the x and y bonds, but can even dominate when an excess is present, which explains the qualitative differences between the scaling found in Fig. \ref{fig:S_lswt}(a) and Fig. \ref{fig:S_lswt}(b-f). In Appendix \ref{App:C} we show that for larger system sizes ($L\sim 120$), also the volume law contribution becomes visible in the geometry studied in Fig. \ref{fig:S_lswt}(a). This heuristic argument is not directly transferable to Fig. \ref{fig:S_lswt}(d), since the magnon entanglement vanishes in the ground state. Instead, in this case we observe a cancellation of the area law contributions coming from the nearest neighbour correlations.

\section{\textbf{Effect of magnon-magnon interactions}}
\begin{figure}
 \includegraphics[width=8.6cm]{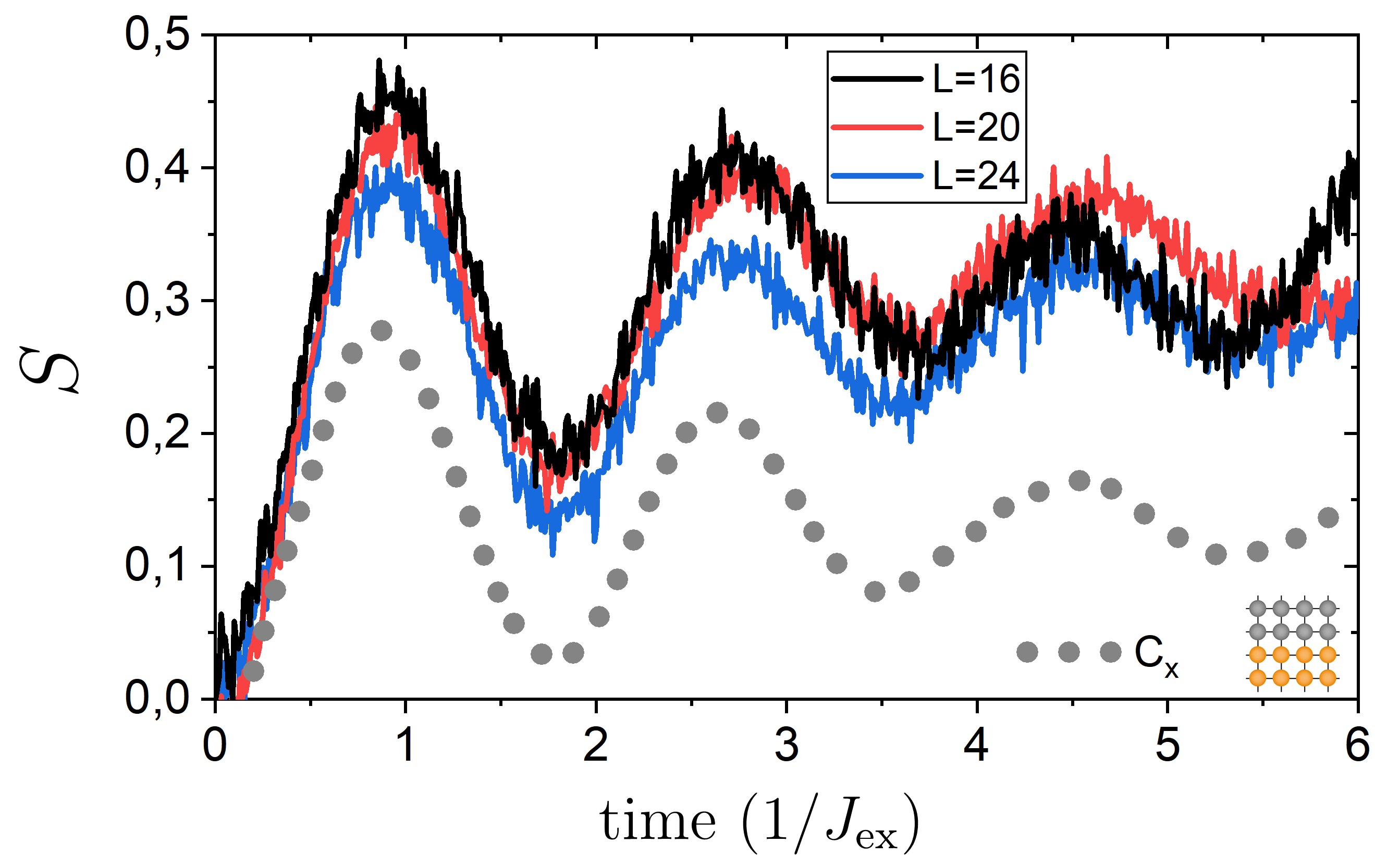} %
    \caption{(Color online) Dynamics of second order Renyi entanglement entropy $S_2(t)$ evaluated with neural-network quantum states for the partition shown at the bottom right, compared with nearest neighbour correlations $C_x(t)\times 16$ (grey dots). $S_2(t)$ shown for $L=16$ ($\alpha=16$, black), $L=20$ ($\alpha=12$, red) and $L=24$ ($\alpha=8$, blue), while $C_x(t)$ for $L=24\, (\alpha=8)$. The Monte Carlo estimations are converged with sampling size.}%
    \label{fig:Fig3}%
  \end{figure} 
  
So far we have considered the dynamics of entanglement only for non-interacting magnon quasiparticles. Although this is generally a good description at long wavelengths, the results above reveal a dominance of short wavelength magnons for which magnon-magnon interactions can significantly influence the dynamics of correlations, especially in the limit of strong quantum fluctuations that we consider here. To investigate if magnon-magnon interactions qualitatively change the entanglement dynamics, we numerically solve Eqs. \eqref{eq:1}-\eqref{eq:2} with the recently proposed \emph{neural-network quantum  states} (NQS) \cite{Carleo}, which were previously adopted to investigate the spreading of correlations in the same setup employed here \cite{Fabiani, Fabiani2}. In this approach, the wavefunction of the system is approximated with a restricted Boltzmann machine: $\psi_M(\sigma) = e^{\sum_i a_i S^z_i}\times \prod^M_{i=1} 2\cosh\big(b_i+\sum_j W_{ij}S^z_j\big)$, where  $\{a_i,b_i,W_{ij}\}$ are the variational parameters, whose number is  $N_{\text{var}} = \alpha \times N^2 + \alpha \times N + N$, with $\alpha=M/N$. These are trained by means of variational Monte Carlo techniques and are time-evolved by employing the time-dependent variational principle  \cite{Carleo 2012}.

To evaluate the entanglement dynamics with NQS, we exploit that $\text{Tr} \rho_{\mathcal{A}}^2$ can be expressed as the expectation value of a \emph{swap} operator between two copies of the system which can be easily computed with Monte Carlo techniques \cite{Hastings}. This allows to efficiently compute the second-order Renyi entropy. The dynamics of $S_2(t)$ is shown in Fig. \ref{fig:Fig3} for system sizes up to $24 \times 24$ spins. Similarly to what is observed within LSWT, the time-scale of the entanglement growth is independent of system size, with the first peak appearing in concomitance with the first peak of nearest neighbour correlations. The rise of entanglement is slightly delayed with respect to LSWT, in close agreement with \cite{Fabiani2}. Such a renormalization is consistent with what is found in spontaneous Raman spectroscopy \cite{Elliot, Canali, Lorenzana} and follows from the appearance of a quasi-bound spin-flip state due to magnon-magnon interactions. Moreover, as can be observed for the largest system, oscillations are strongly damped with time as opposed to LSWT. Although NQS results are limited to a small time-window, relaxation of entanglement may be a signature of thermalization that is triggered by the strong magnon-magnon interactions present in the model.

The dynamics of entanglement shown in Fig. \ref{fig:Fig3} depends weakly on system size, and therefore does not follow the area law found within LSWT for the same geometry. Understanding whether this is an artifact of the numerical approximation (due to finite system size and/or the small $\alpha$ employed, especially for the largest system size) or an effect of magnon-magnon interactions requires more elaborate investigations at even larger systems than considered here. However, we emphasize that the nearest neighbour correlations that dominate the entanglement entropy are already converged.

\section{\textbf{Conclusion}}
We evaluated the entanglement dynamics after a spatially anisotropic quench of the exchange interaction in 2D Heisenberg antiferromagnets. We found that the entanglement is determined by the localized zone-edge magnons and evolves periodically with a frequency solely determined by the exchange energy, independent on the system size. By employing neural-network quantum states, we numerically found that the effect of magnon-magnon interactions is two-fold: at short times they delay the entanglement growth in agreement with the renormalization of the two-magnon mode observed in Raman spectroscopy; at longer times, magnon-magnon interactions lead to a quick relaxation of the oscillations. Whether this is a signature of thermalization is an open question that deserves further investigation. 

We notice that oscillations of the entanglement entropy and/or absence of linear growth have been found in other works in one dimension \cite{Kormos2017, Hodsagi2018,  Alvaredo2019, Alvaredo2020}. For instance, in \cite{Kormos2017} it is shown that meson-like quasiparticles can appear in the post quench dynamics of the Ising chain. In this case, confinement leads to the suppression of light-cone spreading, with correlations and entanglement oscillating at frequencies compatible with the masses of the mesons excited. The present case, however, differs from this scenario as domain walls are absent and there is no suppression of the light cone \cite{Fabiani2}.

Moreover, we found that the dynamics of entanglement is closely related to the dynamics of nearest neighbours correlations. The latter are accessible in state-of-the-art time-resolved Raman scattering experiments \cite{Zhao, Yang, Mazzone2021}, thus providing a direct way to access the dynamics of entanglement in quantum materials.

To conclude, we notice that the quench here employed is relatively weak. Within linear spin wave theory, increasing $\Delta J_\text{ex}$ does not lead to a qualitative difference in the time-evolution as the dynamics remains dominated by the zone-edge magnon pairs and therefore features oscillations at the frequency of the exchange interaction. Moreover, although magnon-magnon interactions renormalize the magnon-pair mode frequency, this qualitative picture is expected to survive even for stronger quenches beyond LSWT. Naturally, distinct physics is expected  for very strong quenches to totally different Hamiltonians, such as quenches to a set of 1D chains with negligible inter-chain interaction. We leave this for future studies. 

\begin{acknowledgments}
This work is part of the Shell-NWO/FOM-initiative ``Computational sciences for energy research” of Shell and Chemical Sciences, Earth and Life Sciences, Physical Sciences, FOM and STW, and received funding from the European Research Council ERC grant agreement No. 856538 (3D-MAGiC). 
\end{acknowledgments}

\appendix

\section{Modified spin wave theory of magnon-pair dynamics}\label{App:A}
In this appendix we give further details concerning the expansion of Eqs. \eqref{eq:1}-\eqref{eq:2} of the main text in terms of magnon operators within the linear spin wave approximation. In particular, we consider the \emph{modified} spin wave theory approach of \cite{Hirsch}. This gets rid of spurious divergences of finite lattices by restoring the sublattice invariance of the Heisenberg Hamiltonian otherwise broken in conventional spin wave treatments. To this end, we add a staggered field to Eq. \eqref{eq:1} of the main text, of the form
\begin{equation}
\hat{\mathcal{H}}_s = - h \sum_{\mathbf{r}_i}e^{\iu \bm{\pi}\cdot \mathbf{r}_i} \hat{S}^z_{\mathbf{r}_i},
\end{equation}
where $e^{\iu \bm{\pi}\cdot \mathbf{r}_i}=+1(-1)$ if $\mathbf{r}$ belongs to sublattice $\mathcal{A}(\mathcal{B})$ of the checkerboard decomposition of the square lattice. This term will be treated in a variational fashion to guarantee that the sublattice invariance is restored, as explained below.  Moreover, in the following we perform a unitary transformation on the spin operators, consisting of a $\pi$ rotation about the $\mathbf{y}$-axis of sublattice $\mathcal{A}$. This yields the transformed operators
\begin{equation} \label{eq:subl_rot}
\hat{\tilde{S}}^z_i = e^{\iu \bm{\pi}\cdot \mathbf{r}_i}\hat{S}^z_i, \hspace{15pt} \hat{\tilde{S}}^x_i = e^{\iu \bm{\pi} \cdot \mathbf{r}_i}\hat{S}^x_i, \hspace{15pt} \hat{\tilde{S}}^y_i = \hat{S}^y_i.
\end{equation}
This transformation allows us to define only one species of boson operators. Here we consider the first-order Holstein-Primakoff transformation
\begin{equation} \label{eq:HP}
\hat{\tilde{S}}_{i}^z = S-\hat{a}^\dagger_{i}\hat{a}_{i}, \hspace{15pt} \hat{\tilde{S}}^+_{i} = \sqrt{2S}\hat{a}_{i},\hspace{15pt}  \hat{\tilde{S}}^-_{i} = \sqrt{2S}\hat{a}^\dagger_{i}.
\end{equation}
It is convenient to work in momentum space, where the Holstein-Primakoff bosons are expressed as
\begin{equation}\label{eq:fourier}
\hat{a}_\mathbf{k} = \frac{1}{\sqrt{N}}\sum_i e^{-\iu{\mathbf{k}}\cdot \mathbf{r}_i}\hat{a}_{i}, \hspace{15pt} \hat{a}_i = \frac{1}{\sqrt{N}}\sum_{\mathbf{k}}e^{\iu\mathbf{k} \cdot \mathbf{r}_i}\hat{a}_{\mathbf{k}},
\end{equation}
where the $i$-sum is over the full lattice, and the $\mathbf{k}$-sum is over the full Brillouin zone. With these transformations we get that $\hat{\mathcal{H}}_T = \hat{\mathcal{H}} + \hat{\mathcal{H}}_s$ becomes (up to constant terms)
\begin{eqnarray}
\hat{\mathcal{H}}_T = \frac{zJ_{\text{ex}}S}{2}\sum_{\mathbf{k}}\bigg[\hat{a}_{\mathbf{k}}^\dagger\hat{a}_{\mathbf{k}} + \hat{a}_{\mathbf{-k}}\hat{a}^\dagger_{\mathbf{-k}} - \gamma_{\mathbf{k}}\Big(\hat{a}_{\mathbf{k}}^\dagger\hat{a}_{-\mathbf{k}}^\dagger+\hat{a}_{\mathbf{k}}\hat{a}_{-\mathbf{k}}\Big)\bigg]\nonumber \\
+ \;  \frac{h}{2}\sum_ \mathbf{k}\big(\hat{a}_{\mathbf{k}}^\dagger\hat{a}_{\mathbf{k}} + \hat{a}_{\mathbf{-k}}\hat{a}^\dagger_{\mathbf{-k}}\big),\hspace{100pt} \nonumber
\end{eqnarray}
where $z=4$ is the coordination number, and $\gamma_{\mathbf{k}} = \frac{1}{z}\sum_{\bm{\delta}}e^{\iu\mathbf{k}\cdot \bm{\delta}}$. This Hamiltonian is diagonalized with a Bogoliubov transformation $\hat{\alpha}_{\mathbf{k}} = \cosh \theta_{\mathbf{k}} \,\hat{a}_{\mathbf{k}} - \sinh \theta_{\mathbf{k}} \,\hat{a}^\dagger_{-\mathbf{k}}$, where $\tanh 2\theta_{\mathbf{k}} = \eta\gamma_{\mathbf{k}}$ and $\eta = \big(1 +\frac{h}{zJ_{\text{ex}}S}\big)^{-1}$, yielding
\begin{equation}\label{eq:lswt_hamiltonian}
\hat{\mathcal{H}}_T = \frac{1}{2}\sum_{\mathbf{k}}\omega_{\mathbf{k}}\Big(\hat{\alpha}_{\mathbf{k}}^\dagger\hat{\alpha}_{\mathbf{k}} + \hat{\alpha}_{-\mathbf{k}}\hat{\alpha}^\dagger_{-\mathbf{k}} \Big), 
\end{equation}
with
\begin{equation}
\omega_{\mathbf{k}}= \frac{zSJ_{\text{ex}}}{\eta}\sqrt{1-(\eta\gamma_{\mathbf{k}})^2}.
\end{equation}
In the spin wave calculation of the main text we employed the Oguchi correction to the single-magnon spectrum \cite{Oguchi}, which is given by the renormalization $\omega_{\mathbf{k}}\rightarrow Z_c \,\omega_{\mathbf{k}}$, where
\begin{equation} \nonumber
    Z_c = 1 + \frac{1}{2S} \frac{1}{N} \sum_{\mathbf{k}} \bigg(1-\sqrt{1-(\eta\gamma_{\mathbf{k}})^2}\bigg) \approx 1.158 .
\end{equation}
This captures the most simple effect of magnon-magnon interactions.

The perturbation $\delta\hat{\mathcal{H}}(t)$ can be written in terms of the same bosons. This basis is convenient because it allows us to express the initial state at $t=0$ as a vacuum state. However, as a consequence the perturbation is not diagonal in this basis. Up to a constant term, we obtain
\begin{eqnarray}
\delta \hat{\mathcal{H}}(t) &= \frac{1}{2}\sum_{\mathbf{k}}\bigg[\delta\omega_{\mathbf{k}}\Big(\hat{\alpha}_{\mathbf{k}}^\dagger\hat{\alpha}_{\mathbf{k}} + \hat{\alpha}_{-\mathbf{k}}\hat{\alpha}^\dagger_{-\mathbf{k}}\Big) \nonumber \\
&+ V_{\mathbf{k}}\Big(\hat{\alpha}_{\mathbf{k}}^\dagger\hat{\alpha}_{-\mathbf{k}}^\dagger + \hat{\alpha}_{\mathbf{k}}\hat{\alpha}_{-\mathbf{k}}\Big)\bigg],
\end{eqnarray}
with
\begin{eqnarray}
\delta \omega_{\mathbf{k}} &=& zS\Delta J_{\text{ex}}(t) \frac{1-\eta\xi_{\mathbf{k}}\gamma_{\mathbf{k}}}{\sqrt{1-(\eta\gamma_{\mathbf{k}})^2}}, \nonumber \\
V_{\mathbf{k}} &=& -zS\Delta J_{\text{ex}}(t) \frac{\xi_{\mathbf{k}}-\eta\gamma_{\mathbf{k}} }{\sqrt{1-(\eta\gamma_{\mathbf{k}})^2}} ,
\end{eqnarray}
where we defined $\xi_{\mathbf{k}} = \frac{1}{2}\sum_{\bm{\delta}}(\mathbf{e}\cdot \bm{\delta})^2 e^{i\mathbf{k}\cdot \bm{\delta}}$.
The value of $h$ is adjusted such that the staggered magnetization $m_z = \sum_{\mathbf{r}_i}e^{\iu \bm{\pi}\cdot \mathbf{r}_i} \braket{\hat{S}^z_{\mathbf{r}_i}}=0$ in order to restore the sublattice invariance of Eqs. \eqref{eq:1}-\eqref{eq:2}. This is obtained by requiring
\begin{equation}\label{eq:mod_condition}
\frac{1}{2N}\sum_{\mathbf{k}}\frac{2\braket{\hat{K}^z_\mathbf{k}}-\eta\gamma_\mathbf{k}\braket{\hat{K}^+_\mathbf{k}+\hat{K}^-_\mathbf{k}}}{\sqrt{1-(\eta\gamma_\mathbf{k})^2}} -\frac{1}{2}= S
\end{equation} 
where we have defined the magnon-pair operators \cite{Bossini 2019}
\begin{eqnarray}
\hat{K}^z_{\mathbf{k}} = \frac{1}{2}\Big(\hat{\alpha}_{\mathbf{k}}^\dagger\hat{\alpha}_{\mathbf{k}} + \hat{\alpha}_{-\mathbf{k}}\hat{\alpha}^\dagger_{-\mathbf{k}}\Big),\nonumber \\
\hat{K}^+_{\mathbf{k}} = \hat{\alpha}_{\mathbf{k}}^\dagger\hat{\alpha}_{-\mathbf{k}}^\dagger, \hspace{15pt} \hat{K}^-_{\mathbf{k}} = \hat{\alpha}_{\mathbf{k}}\hat{\alpha}_{-\mathbf{k}}. \nonumber
\end{eqnarray}
Note that $\braket{\hat{K}^z_\mathbf{k}}$ and $\braket{\hat{K}^\pm_\mathbf{k}}$ are time-dependent and therefore $\eta$ acquires a time-dependence through Eq. \eqref{eq:mod_condition}, which has to be solved self-consistently
at each step of the time-evolution.

The dynamics of $\braket{\hat{K}^z_\mathbf{k}}$ and $\braket{\hat{K}^\pm_\mathbf{k}}$ is found by numerically solving the equations of motion
\begin{eqnarray}
\frac{d\big\langle\hat{K}^z_{\mathbf{k}}\big\rangle}{dt} &=& \iu V_{\mathbf{k}}\Big[\big\langle\hat{K}^-_{\mathbf{k}}\big\rangle-\big\langle\hat{K}^+_{\mathbf{k}}\big\rangle\Big], \\
\frac{d\big\langle\hat{K}^{\pm}_{\mathbf{k}}\big\rangle}{dt} &=& \pm 2 \iu\Big[(\omega_{\mathbf{k}}+\delta\omega_{\mathbf{k}})\big\langle\hat{K}^{\pm}_{\mathbf{k}}\big\rangle + V_{\mathbf{k}}\big\langle\hat{K}^z_{\mathbf{k}}\big\rangle\Big],
\end{eqnarray}
together with Eq. \eqref{eq:mod_condition} and the initial conditions $\braket{\hat{K}^{z}_{\mathbf{k}}}=\frac{1}{2}$ and $\braket{\hat{K}^{\pm}_{\mathbf{k}}}=0$. The solution of these equations are qualitatively similar to the corresponding solutions for $h=0$ found in \cite{Fabiani2}. The dynamics of correlations is given by $C(\mathbf{R},t)=\frac{2S}{N}\sum_{\mathbf{k}}C_\mathbf{k}$, where
\begin{equation}\label{eq:C_LSWT}
C_\mathbf{k} = \frac{\cos(\mathbf{k}\cdot\mathbf{R})}{\sqrt{1-(\eta\gamma_{\mathbf{k}})^2}}\bigg[ \big\langle \hat{K}^z_{\mathbf{k}}\big\rangle + \frac{\eta\gamma_{\mathbf{k}}}{2}\Big( \big\langle \hat{K}^+_{\mathbf{k}}\big\rangle+ \big\langle \hat{K}^-_{\mathbf{k}}\big\rangle\Big)-\frac{1}{2}\bigg],
\end{equation}
for $\mathbf{R}$ connecting spins in the same sublattice; a similar expression can be obtained when $\mathbf{R}$ connects different sublattice spins.

\section{\textbf{II. Two-level system approximation}}\label{App:B}
The calculations of the main text based on the two-level system approximation are here reviewed and complemented with further details. This approximation assumes that the system can be regarded as a collection of independent two-level systems, each of them consisting of a superposition of the vacuum state and a state describing a pair of magnons with wavevector $\{\mathbf{k},-\mathbf{k}
\}$. This yields the following ansatz for the wavefunction: $\ket{\psi(t)} = \prod_{\mathbf{k}}\ket{\psi_{\mathbf{k}}(t)}$, where 
\begin{equation}\label{eq:2L}
\ket{\psi_{\mathbf{k}}(t)} = c_{\mathbf{k}}(t) \ket{0_{\mathbf{k}} 0_{-\mathbf{k}}} + d_{\mathbf{k}}(t) \ket{1_{\mathbf{k}} 1_{-\mathbf{k}}}.
\end{equation}
All the time-dependence of the wavefunction is captured in the coefficients $c_\mathbf{k}(t)$ and $d_\mathbf{k}(t)$. At $t=0$ the system is in the ground state of Eq. \eqref{eq:lswt_hamiltonian}, which is the vacuum state of the magnon operators $\alpha_\mathbf{k}$, and therefore $c_\mathbf{k}(0)=1$ and $d_\mathbf{k}(0)=0$. The time-evolution of each $\ket{\psi_{\mathbf{k}}(t)}$ can be found by projecting the Schroedinger equation onto the states $\ket{0_{\mathbf{k}} 0_{-\mathbf{k}}}$ and $\ket{1_{\mathbf{k}} 1_{-\mathbf{k}}}$. This yields the following set of equations
\begin{eqnarray}
i\frac{d}{dt}c_\mathbf{k}(t) &=& V_\mathbf{k} d_\mathbf{k}(t), \\
i\frac{d}{dt}d_\mathbf{k}(t) &=& V_\mathbf{k} c_\mathbf{k}(t) + 2(\omega_\mathbf{k} + \delta\omega_\mathbf{k})d_\mathbf{k}(t).  
\end{eqnarray}
These coupled equations can be solved exactly, yielding
\begin{eqnarray}
c_\mathbf{k}(t) &=& \Big[ \cos a_\mathbf{k} t + i\frac{\omega_\mathbf{k}+\delta\omega_\mathbf{k}}{a_\mathbf{k}}\sin a_\mathbf{k}t\Big]e^{-i(\omega_\mathbf{k}+\delta\omega_\mathbf{k})t}, \nonumber \\
d_\mathbf{k}(t) &=& -i\frac{V_\mathbf{k}}{a_\mathbf{k}}e^{-i(\omega_\mathbf{k}+\delta\omega_\mathbf{k})t}\sin a_\mathbf{k} t,\nonumber
\end{eqnarray}
with $a_\mathbf{k} = \sqrt{(\omega_\mathbf{k}+\delta\omega_\mathbf{k})^2+V_\mathbf{k}^2}$.

The form of the wavefunction Eq. \eqref{eq:2L} suggests that the magnons of each two-level system are highly entangled. Following the main text, we calculate the second order Renyi entropy between the mode with wavevector $+\mathbf{k}$ and the mode with wavevector $-\mathbf{k}$. To this purpose, we introduce the density matrix $\rho_\mathbf{k} = \ket{\psi_\mathbf{k}}\bra{\psi_\mathbf{k}}$; then, the reduced density matrix $\tilde{\rho}_\mathbf{k}$ of the mode with wavevector $-\mathbf{k}$ can be obtained by tracing out the first boson (with wavevector $+\mathbf{k}$) from $\rho_\mathbf{k}$
\begin{eqnarray}
\tilde{\rho}_\mathbf{k} &=& \braket{0_\mathbf{k}|\rho_\mathbf{k}|0_\mathbf{k}} + \braket{1_\mathbf{k}|\rho_\mathbf{k}|1_\mathbf{k}}\nonumber \\
&=& |c_\mathbf{k}(t)|^2 \ket{0_{-\mathbf{k}}}\bra{0_{-\mathbf{k}}} + |d_\mathbf{k}(t)|^2 \ket{1_{-\mathbf{k}}}\bra{1_{-\mathbf{k}}}\nonumber \\
&=& \begin{pmatrix}
|c_\mathbf{k}(t)|^2 & 0\\
0 & |d_\mathbf{k}(t)|^2
\end{pmatrix}.
\end{eqnarray}
Note that alternatively we could trace out the boson with wavevector $-\mathbf{k}$; this would yield the same reduced density matrix.
Finally, we get the second order Renyi entropy 
\begin{equation}
S^{\text{bog}}_{\text{2L}} = -\ln\big(\text{Tr}\tilde{\rho}^2_\mathbf{k}\big) = -\ln\big(|c_\mathbf{k}(t)|^4 + |d_\mathbf{k}(t)|^4\big).
\end{equation}

\section{\textbf{III. Additional results on correlation and entanglement dynamics}}
\label{App:C}
\begin{figure}
   {\includegraphics[width=8.6cm]{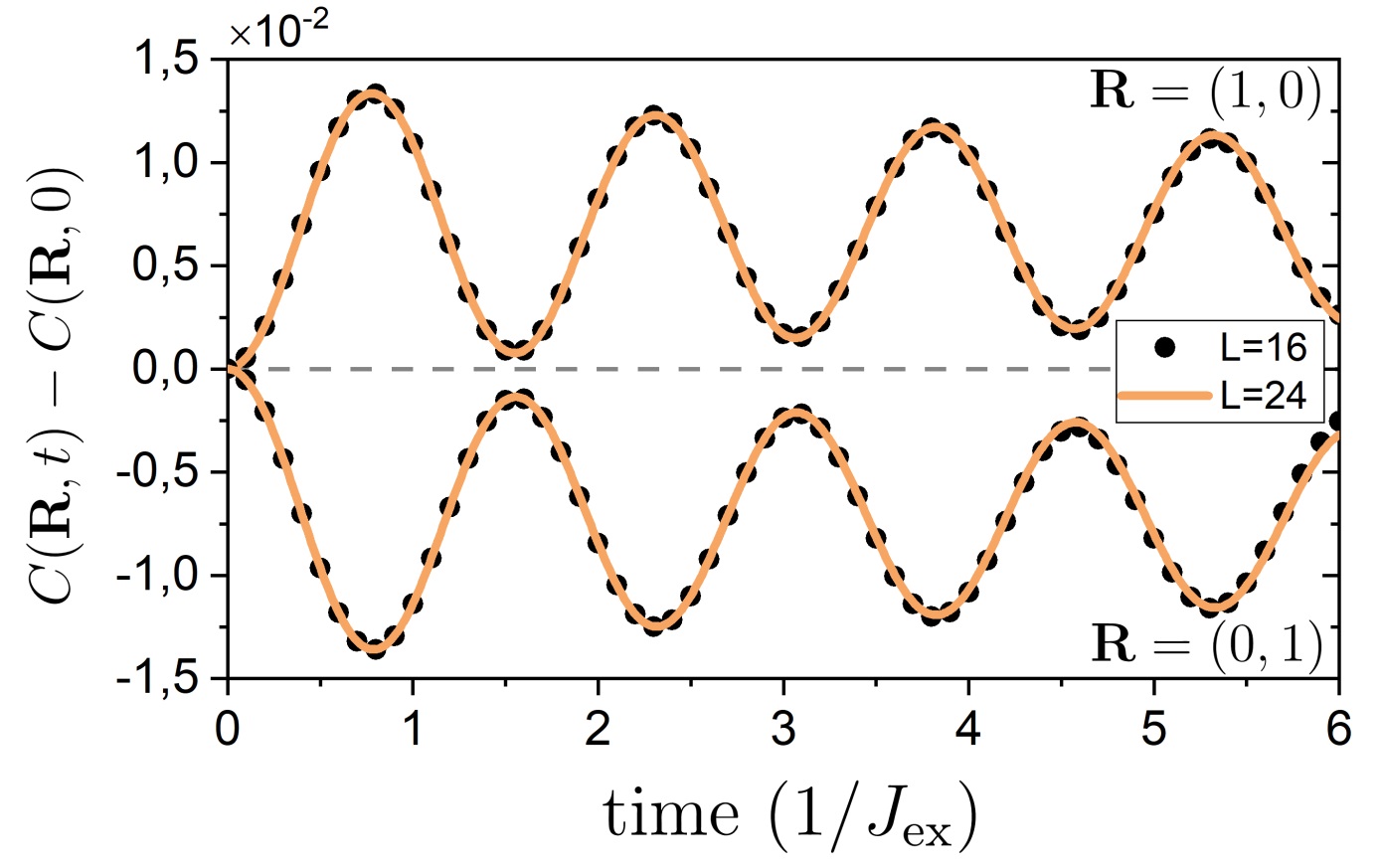} }%
    \caption{(Color online) Dynamics of nearest neighbour correlations with $\mathbf{R}=(1,0)$ and $\mathbf{R}=(0,1)$ for $L=24$ (solid orange line) and $L=16$ (black dots) evaluated within the linear spin wave approximation.}%
    \label{fig:Fig1App}%
  \end{figure} 
  
Here we provide additional results about the correlation dynamics and the scaling of entanglement with system size in the linear spin wave approximation.
The dynamics of nearest neighbours correlations evaluated with Eq. \eqref{eq:C_LSWT} is shown in Fig \ref{fig:Fig1App} for $L=16,\, 24$ and $\mathbf{R}=[1,0]$, $\mathbf{R}=[0,1]$, revealing that correlations along the x and y direction of the lattice oscillate out of phase as expected from the symmetry of the perturbation \cite{Fabiani2}. Note that the nearest neighbours correlations have converged with system size already for the small systems considered here and therefore no deviations from $L=24$ are found at larger $L$. Differently, the entanglement dynamics has a strong dependence with system size. In the main text it was observed that if one considers a partition of the lattice with exceeding number of nearest neighbour correlations along one direction, the dynamics of the entanglement entropy in the Holstein-Primakoff basis is mainly determined by such nearest neighbour correlations, resulting in a area law scaling of the entanglement dynamics. To better understand this, let us write the reduced density matrix $\rho_{\mathcal{A}(\mathcal{B})}$ in terms of spin operators
\begin{equation}\label{eq:rho}
\rho_{\mathcal{A}(\mathcal{B})} = 2^n\sum_{\mu, \dots ,\nu=0}^3 \braket{\hat{S}_{i_1}^\mu \cdots \hat{S}^\nu_{i_n}}\hat{S}_{i_1}^\mu  \otimes \cdots \otimes \hat{S}^\nu_{i_n},
\end{equation}
where $n$ is the number of spins in $\mathcal{A}(\mathcal{B})$ and  $\hat{S}^0_{i}=\frac{1}{2}\mathbb{I}$ and $\mu=\{1,2,3\}$ correspond to $\{x,y,z\}$. Recalling that the second-order Renyi entropy is defined as $S_2(\rho_{\mathcal{A}(\mathcal{B})})\equiv S_2=-\text{Tr}\rho_{\mathcal{A}(\mathcal{B})}^2$, it is possible to show that the entanglement entropy can be expressed in terms of  the sum of all the possible correlations within $\mathcal{A}(\mathcal{B})$, which enter in $S_2$ with a square. Since nearest neighbouring correlations are much larger than the other correlations \cite{Fabiani2}, the leading order contributions to the entanglement entropy can be written as
\begin{equation}\label{eq:S_2_app}
S_2 \sim -\ln{\Big[a_0 + a_1\sum_{i \in \mathcal{A}(\mathcal{B})} \braket{\hat{S}_i \cdot \hat{S}_{i+\delta_x}}^2 + \braket{\hat{S}_i \cdot \hat{S}_{i+\delta_y}}^2\Big]},
\end{equation}
for some constants $a_0$ and $a_1$. Since at leading order in $\Delta J_{\text{ex}}$ it holds that $\braket{\hat{S}_i \cdot \hat{S}_{i+\delta_{x(y)}}}=\text{const}\,+(-)\,\Delta J_{\text{ex}}s(t)$, for some oscillating function $s(t)$, then at leading order also $\braket{\hat{S}_i \cdot \hat{S}_{i+\delta_x}}^2$ and $\braket{\hat{S}_i \cdot \hat{S}_{i+\delta_y}}^2$ oscillate out of phase. Therefore, it follows from Eq. \eqref{eq:S_2_app} that if there is an equal amount of correlations along $\hat{x}$ and $\hat{y}$, then the leading order contribution to $S_2$ vanishes. To the contrary, when there is an exceeding number of correlations along one direction, let say $\hat{x}$, then $S_2\sim -\ln{\Big[a_0 + a_1\sum_{i \in \partial\mathcal{A}(\mathcal{B})} \braket{\hat{S}_i \cdot \hat{S}_{i+\delta_x}}^2 \Big]}$, where $\partial \mathcal{A}(\mathcal{B})$ is the boundary of the $\mathcal{A}(\mathcal{B)}$ partition. Hence, the symmetry of the perturbation and the dominance of nearest neighbor correlations explain the qualitative difference in the dynamics of the entanglement entropy between the partition considered in Fig. \ref{fig:S_lswt}(a) of the main text, which has an exceeding number of bonds along the $x$-direction, and the other partitions, where the number of $x$ and $y$ bonds is balanced. In addition, for the other partitions a volume law scaling was found, which we ascribe to next-to-leading order terms, as well as to the correlations neglected in Eq. \eqref{eq:S_2_app}.

\begin{figure}
   {\includegraphics[width=8.6cm]{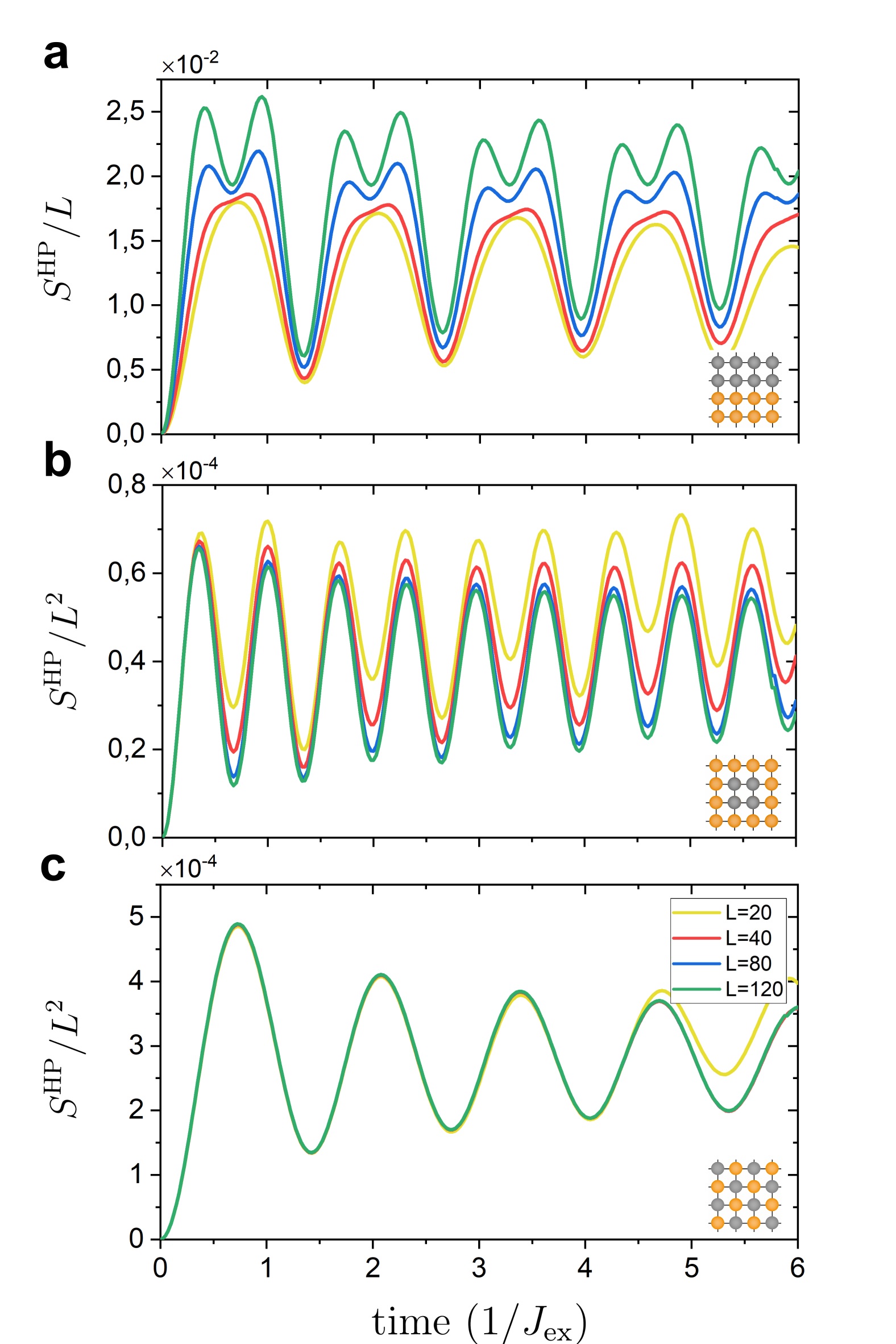} }%
    \caption{(Color online) (a)-(c) Dynamics of Renyi entanglement entropy in the Holstein-Primakoff basis for $L=20,\, 40,\, 80,\, 120$ for the partitioning shown in the inset of each plot.}%
    \label{fig:Fig2App}%
  \end{figure} 

Even though nearest neighbor correlations dominate and are converged already for small $L$, the contribution of area law and volume law entanglement strongly depends on system size, which suggest that in large systems  the volume law contributions will become visible even in the partition with an exceeding number of bonds along one direction. In the following we show that this is indeed the case by reporting additional results concerning the finite-size scaling of the entanglement within LSWT. In Fig \ref{fig:Fig2App} we plot the entanglement dynamics in the Holstein-Primakoff basis for the same partitioning shown in the main text, but for system sizes up to $L=120$. By comparing Fig. \ref{fig:S_lswt}(a) of the main text and Fig. \ref{fig:Fig2App}(a) we observe that while for small systems the entanglement neatly follows an area law scaling, when increasing system size a volume law contribution builds up. The effect of this term comprises an additional oscillation with double the frequency of those seen at small system sizes and similar to what is observed for the other partitions where the area law contribution is absent. Indeed, in these other partitions a volume law scaling is maintained also at large system sizes, as can be observed in Figs. \ref{fig:Fig2App}(b)-(c). We conclude by noticing that the appearance of such double-frequency oscillations is due to the structure of the matrix $Q=X \cdot P $ introduced in the main text. Both $X$ and $P$ are dominated by nearest neighbour correlations, which oscillate with the characteristic two-magnon frequency $\omega_{\text{2M}}$. As such, $Q$ contains both terms oscillating at $\omega_{\text{2M}}$ (at leading order in $\Delta J_{\text{ex}}$) and $2\times\omega_{\text{2M}}$ (at next-to-leading order in $\Delta J_{\text{ex}}$), as also follows from the discussion below Eq. \eqref{eq:S_2_app}. We numerically verified that in the Bogoliubov basis, the leading order terms oscillating in $Q_{ij}(t)=\sum_k X_{ik}(t)P_{kj}(t)$ at $\omega_{\text{2M}}$ cancel out due to the vanishing of correlations at $t=0$, which explains the difference between Fig. \ref{fig:S_lswt}(a) and Fig. \ref{fig:S_lswt}(d) of the main text.

\end{document}